\documentclass[10pt]{article} 
\usepackage[preprint]{rlc}

\usepackage{amssymb}            
\usepackage{mathtools}          
\usepackage{mathrsfs}           
\mathtoolsset{showonlyrefs}     
\usepackage{graphicx}           
\usepackage{subcaption}         
\usepackage[space]{grffile}     
\usepackage{url}                

\usepackage{amsmath} 
\usepackage{algorithm}
\usepackage{algpseudocode}
\usepackage{layout}
\usepackage{hyperref}
\usepackage{amsmath} 
\usepackage{xspace}
\usepackage{bm}
\usepackage[table]{xcolor}

\definecolor{softred}{RGB}{200,50,50}
\definecolor{softblue}{RGB}{40,90,160}

\newcommand{\our}{\textsc{SocialDriveGen}\xspace}

\usepackage{booktabs}
\title{ SocialDriveGen:~Generating~Diverse~Traffic~Scenarios with  Controllable Social Interactions
}

\author{\textbf{Jiaguo Tian}$^{1}$, \textbf{Zhengbang Zhu}$^{1}$, \textbf{Shenyu Zhang}$^{1}$, \textbf{Li Xu}$^{2}$, \textbf{Bo Zheng}$^{2}$, \\
  \textbf{Xu Liu}$^{2}$, \textbf{Weiji Peng}$^{2}$, \textbf{Shizeng Yao}$^{2}$, \textbf{Weinan Zhang}$^{1}$ \\
  \\ 
  $^{1}$Shanghai Jiao Tong University, \,
  $^{2}$Chongqing Changan Automobile Co. Ltd \\
}

\begin{document}

\maketitle
\thispagestyle{empty}
\pagestyle{empty}

\begin{abstract}
The generation of realistic and diverse traffic scenarios in simulation is essential for developing and evaluating autonomous driving systems. However, most simulation frameworks rely on rule-based or simplified models for scene generation, which lack the fidelity and diversity needed to represent real-world driving. While recent advances in generative modeling produce more realistic and context-aware traffic interactions, they often overlook how social preferences influence driving behavior. SocialDriveGen addresses this gap through a hierarchical framework that integrates semantic reasoning and social preference modeling with generative trajectory synthesis. By modeling egoism and altruism as complementary social dimensions, our framework enables controllable diversity in driver personalities and interaction styles. Experiments on the Argoverse 2 dataset show that SocialDriveGen generates diverse, high-fidelity traffic scenarios spanning cooperative to adversarial behaviors, significantly enhancing policy robustness and generalization to rare or high-risk situations.

\end{abstract}

\section{Introduction}

The development of safe and reliable autonomous driving systems\citep{caesar2021nuplan, ding2023survey, hu2023planning, cui2024drive} has become a central focus in modern transportation research, aiming to enhance mobility, reduce accidents, and improve driving efficiency through advanced algorithmic design and large scale testing. Real world testing alone is infeasible as demonstrating human level safety statistically would require billions of kilometers and even such testing cannot exhaust the long tail of rare, safety critical ``edge cases". Simulation has therefore become an indispensable paradigm for autonomous driving validation. It enables scalable and efficient evaluation of driving algorithms, providing a repeatable and controllable environment to test rare, high risk interactions such as aggressive cut-ins or unpredictable pedestrians, which are essential for assessing real world robustness.

\begin{figure}[htbp]
    \centering
    \includegraphics[width=1.0\textwidth]{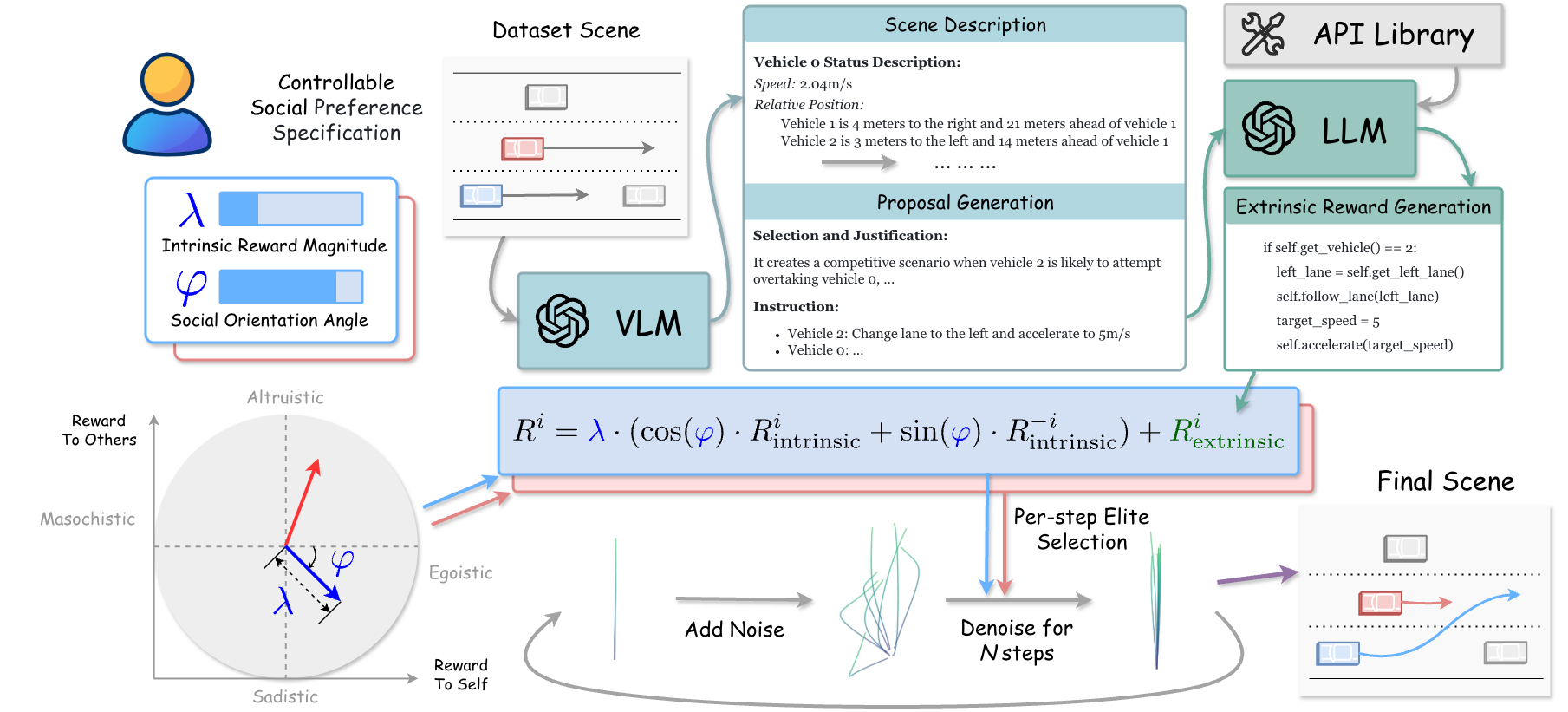}
    \caption{Overview of \our. The framework leverages VLMs and LLMs to construct social rewards that guide a multi-agent diffusion process.}
    \vspace{-8pt}
    \label{fig:social-drive-gen}
\end{figure}

As simulation technology continues to evolve, the main focus of simulation fidelity has shifted from replicating only the physical world toward capturing the intricacies of human behavior and social interaction.  Early simulation research~\citep{feng2022trafficgen, ding2024realgen} emphasized accurate vehicle dynamics, sensor modeling, and photorealistic environments to ensure physical credibility. However, as autonomous driving systems increasingly engage in decision making and interaction with other agents, the need for behavioral and social realism has become crucial~\citep{zhan2019interaction}. Simulated social vehicles must display diverse strategies resembling human drivers, including cooperation, competition, negotiation, and occasional irrationality, to adequately assess the ability of modern driving policies~\citep{schwarting2019social, jin2024surrealdriver}. Traditional rule based traffic models~\citep{treiber2000congested} are computationally efficient but inherently limited in diversity, often producing overly deterministic and predictable behaviors~\citep{suo2021trafficsim}. 
Psychology inspired approaches using Social Value Orientation (SVO)~\citep{schwarting2019social} improve behavioral realism by modeling how agents balance self and others' rewards. While effective for predicting cooperative and competitive driving, current applications~\citep{dai2023socially} focus on limited and simplified scenarios. Extending SVO to dynamic, multi-agent contexts remains a key challenge for realistic social behavior modeling.

Recent progress in generative models provides new opportunities to address these challenges. Vision Language Models (VLMs) and Large Language Models (LLMs)~\citep{openai2023gpt4, liu2023visual} demonstrate strong reasoning and multimodal understanding, allowing them to serve as semantic planners that interpret driving scenes and propose high level interaction intents~\citep{xu2024drivegpt4, fu2024drive}. At the same time, diffusion models have become the leading approach for producing temporally coherent and physically consistent multi-agent trajectories. Integrating these paradigms by connecting high level intent reasoning from language models with trajectory generation from diffusion models has shown promising results~\citep{zhang2025drivegen}. However, current studies rarely incorporate social value modeling, which limits the diversity and social awareness of the generated interactions.

To overcome these limitations, we propose SocialDriveGen, a hierarchical framework that unites semantic reasoning, social psychology, and generative modeling to create controllable and socially realistic traffic scenarios. A VLM first analyzes scene context and proposes meaningful interactions between vehicles, which are then interpreted by an LLM into practical reward functions that define interpretable objectives for generation. Building upon this, a socially aware reward formulation explicitly represents egoism and altruism, offering fine grained control over driver personalities across the behavioral spectrum. Finally, a guided multi-agent diffusion model jointly synthesizes trajectories for all vehicles, where each denoising step is directed by an evolutionary strategy that gradually steers generation toward high reward samples. Together, these components enable the creation of diverse and realistic scenarios that better capture the social nature of driving.

Our key contributions are summarized as follows:
\begin{itemize}
    \item We introduce SocialDriveGen, a unified framework that combines semantic reasoning and generative modeling to produce controllable and socially realistic driving scenarios.
    \item We propose a two dimensional social reward formulation based on egoism and altruism, enabling precise control over diverse driving personalities and social interactions.
    \item We develop a guided multi-agent diffusion process that jointly synthesizes trajectories for all vehicles, achieving socially complex and high fidelity scenarios that improve policy robustness and diversity.
\end{itemize}

\section{Method}

%
In this section, we propose a framework for social-aware driving scenario generation. First, we introduce VLMs for high level scenario analysis and social interaction proposal. Second, we propose a reward integration module to judge the reward of a vehicle, which is also capable of depicting the different social personality of drivers. Third, we guide a multi-agent diffusion model for joint trajectory generation. 


\subsection{Scenario Analysis and Proposal Generation with VLM}

We leverage Visual-Language Models (VLMs) to identify and generate adverserial interactive scenario proposals from real-world driving datasets. The process is executed in two distinct stages:

\textbf{Semantic Scene Description Stage} We define the state of a single vehicle $i$ at time $t$ be denoted by a vector $\mathbf{s_t^i}=[p_t^i,v_t^i,\theta_t^i]$, representing its position, velocity, and heading. A vehicle's trajectory is thus a sequence of states, $\tau_i=\{s_1^i,s_2^i,\dots ,s_{T_s}^i\}$, where $T_s$ is the total duration of the traffic. The entire scenario, extracted from a real-world dataset, is represented as a tuple $\mathcal{S}=(\cup_{i=1}^N \tau_i,\mathcal{M})$, where $N$ is the total number of vehicles and $\mathcal{M}$ denotes the visual map data. By analyzing the state $S$, the VLM generates a natural language description $D_{\text{scene}}= \text{VLM}(\mathcal{S})$ that captures the scene's dynamic features including relative spatial positions, relative speed and infers underlying social intentions. 

\textbf{Adversarial Proposal Generation Stage} Subsequently, the VLM acts as a strategic proposer. Conditioned on the scene description $D_{\text{scene}}$ and visual map data $\mathcal{M}$, it selects a pair of vehicles $\{v_i,v_j\}$ from the set of all vehicles with the highest potential for social interaction. It is prompted to generate specific, human-like adversarial proposals, $\mathcal{P}_{i, j} = \{\mathcal{A}_i, \mathcal{A}_j\}$, for these two vehicles. Here, $\mathcal{A}_i$ and $\mathcal{A}_j$  are natural language instructions that specify high level intentions corresponding to vehicles $v_i$ and $v_j$. For \our,  $\mathcal{P}_{i, j}$ serve as critical conditioning signals for the downstream trajectory generation process, ensuring that the final generated scenarios are highly interactive and diverse.

\subsection{Social-aware Reward Formulation}

To achieve the simulation of varied driving scenarios, we design a social-aware reward function inspired by established social psychological theories and their extensions. This function evaluates driving behaviors and guides the downstream generative model to generate scenarios across a broad spectrum of game-theoretic interactions, from cooperative to highly adversarial.

\subsubsection{Decomposition of Social Behavior} 

To achieve a more nuanced representation of driving behavior, we decompose social motivation along two critical axes, addressing key limitations in recent literature. 

First, prior works~\citep{schwarting2019social, dai2023socially} often represent Social Value Orientation (SVO) on a 1D spectrum. These approaches conflate distinct motivations—such as a reckless agent (low concern for self, low concern for others) and a purely selfish agent (high concern for self, low concern for others)—into a single ambiguous value. More fundamentally, these models often incorrectly assume the scope of social preference. They implicitly define the SVO of an agent $i$ as applying to the entirety of another's reward, $R_{-i}$, which conflates the other's intrinsic values (e.g., safety, comfort) with their extrinsic, task-specific goals, $R_{-i}^{\text{extrinsic}}$. This is unrealistic, since the extrinsic motivation or task-goal of another vehicle is generally unobservable to the driver. Therefore, a driver's social preference (e.g., altruism) can realistically only apply to the other's inferred intrinsic values, such as their safety and comfort, not their specific, hidden goals.

\subsubsection{The Integrated Reward Function} 

Building on the dual decomposition from the previous section, we now formalize the integrated reward function for agent $i$. The total reward $R_i$ is composed of two primary components: a social-aware intrinsic term and a task-oriented extrinsic term.

These components are balanced by a intrinsic reward magnitude $\lambda$, which controls the agent's tendency to pursue extrinsic goals versus intrinsic values. The intrinsic term itself is governed by a social orientation angle $\phi$, which sets the agent's SVO angle. This is formulated as:

\begin{equation} \label{eq:social_reward}
R_i = \lambda \cdot ( \cos(\phi) \cdot R^{i}_{\text{intrinsic}} + \sin(\phi) \cdot R^{-i}_{\text{intrinsic}}) + R^{i}_{\text{extrinsic}}
\end{equation}

The \textbf{extrinsic reward} ($R_{\text{extrinsic}}$) is dynamically generated by the LLM to be task-oriented. Based on the natural language proposal $\mathcal{P}_{i,j}$ from the previous stage generated by the VLM, the LLM parses the specific objective (e.g., "successfully complete a lane change") and invokes a predefined reward API to instantiate a concrete reward function. This design makes our framework highly extensible, allowing it to adapt to diverse adversarial proposals. In contrast, the \textbf{intrinsic rewards} ($R_{\text{intrinsic}}$) are internally motivated, reflecting the inherent preferences of the driver. $R_{\text{intrinsic}}$ accounts for a driver's personal comfort, safety, and adherence to standard driving rules, including lane keeping, speed maintenance, and heading consistency.

\subsection{Multi-vehicle Trajectory Generation} 

The downstream generation module of \our generates multi-agent trajectories that maximize the social-aware reward functions defined in the previous section. Our approach adapts the Evolutionary Strategy (ES)-guided diffusion model~\citep{yang2024diffusion} to a multi-agent setting. Unlike prior work that typically applies reward guidance only to the final denoised output, our method integrates this guidance throughout the entire denoising process.

\subsubsection{Denoising Process with Step-wise Guidance} Our key innovation is to integrate a reward-based evolutionary search into each iterative denoising step, progressively refining trajectories toward high-reward solutions.

Given a trained diffusion model, a standard reverse diffusion process for a single step can be expressed as: 
\begin{equation} p_\theta(\mathbf{x}_{t-1}|\mathbf{x}_t) = \mathcal{N}(\mathbf{x}_{t-1}; \boldsymbol{\mu}_\theta(\mathbf{x}_t, t), \boldsymbol{\Sigma}_\theta(\mathbf{x}_t, t)) \end{equation}
 where $x_t$ is a noisy sample at timestep $t$, and the model $\theta$ predicts the mean $\mu$ and covariance $\Sigma$ to generate a less noisy sample $\mathbf{x}_{t-1}$.

Our method introduces a gradient-free evolutionary guidance loop that leverages our reward function. This process begins with an initial population of $M$ trajectories. At each search step $k$, the entire population is renoised to promote exploration. The core of our guidance then occurs within the inner denoising loop. At each denoising step $t$, we first evaluate the entire noisy population $\mathbf{X}_t$. For each noisy sample $\mathbf{x}_t^{(i)}$, we predict its corresponding clean trajectory $\mathbf{x}_0^{(i)}$ and evaluate its reward, $R(\mathbf{x}_0^{(i)})$. This reward is then assigned back to  $\mathbf{x}_t^{(i)}$.

A reward-weighted "elite" distribution $q$ is then computed over this noisy population $\mathbf{X}_t$, formally defined as:
\begin{equation}
q(\mathbf{x}) = \frac{\exp(\tau \cdot R(\mathbf{x}))}{\sum_{j=1}^{M} \exp(\tau \cdot R(\mathbf{x}_t^{(j)}))}
\end{equation}
where $\tau$ is the temperature parameter. This distribution $q$ is then used to sample a population of ``elites" from $\mathbf{X}_t$, which forms the basis for the subsequent denoising step $t-1$. This step-wise selection loop continuously guides the generation process, progressively refining the trajectories toward high-reward solutions.

Crucially, the temperature parameter $\tau$ is not fixed. We employ an annealing schedule for $\tau$, increasing its value as the denoising process progresses. This strategy enables an exploratory search in the early stages when noise is high, and transitions to an exploitative search in the later stages when the trajectories are more refined. This approach balances exploration (diversity) with exploitation (high-reward solutions).

\subsubsection{Joint Multi-Agent Trajectory Generation}
We extend the single-agent diffusion framework to a multi-agent setting by modeling the joint trajectory distribution of all vehicles. Unlike individual generation, this approach explicitly captures inter-agent dependencies. By guiding this joint process with a social-aware reward function, we optimize the entire interactive scenario to align with high-level adversarial proposals. This ensures that collective behaviors are not only cohesive and physically plausible but also conform to the desired game-theoretic outcomes.

\section{Experiment} In this section, We conduct experiments to evaluate \our, focusing on two key objectives: to validate our framework's ability to generate diverse, controllable, and challenging scenarios, and to demonstrate the effectiveness of our core components: the two-stage VLM planner and the gradient-free guidance mechanism. All experiments are conducted on the Argoverse 2 dataset~\citep{wilson2023argoverse}.We leverage these real-world recordings as a foundation to generate scenarios enriched with complex social interactions.

We evaluate performance using key interaction metrics. To evaluate the quality of generated scenarios, we compute Engagement Ratio.  The Engagement Ratio is calculated by setting a threshold for Time-to-Collision (TTC), defined as the estimated time until a collision between vehicles based on their current trajectories and speeds. Scenarios with a maximum TTC exceeding this threshold are classified as engagement scenarios. We also measure the maximum relative velocity between vehicles to quantify interaction intensity. Furthermore, we compute the maximum acceleration to capture the aggressiveness and non-cooperative nature of the driving behaviors. To evaluate task completion, we report the Extrinsic Reward regarding the vehicle's trajectory, which reflects the specific driving objectives.

\begin{figure}[htbp]
    \centering
    \includegraphics[width=1.0\textwidth]{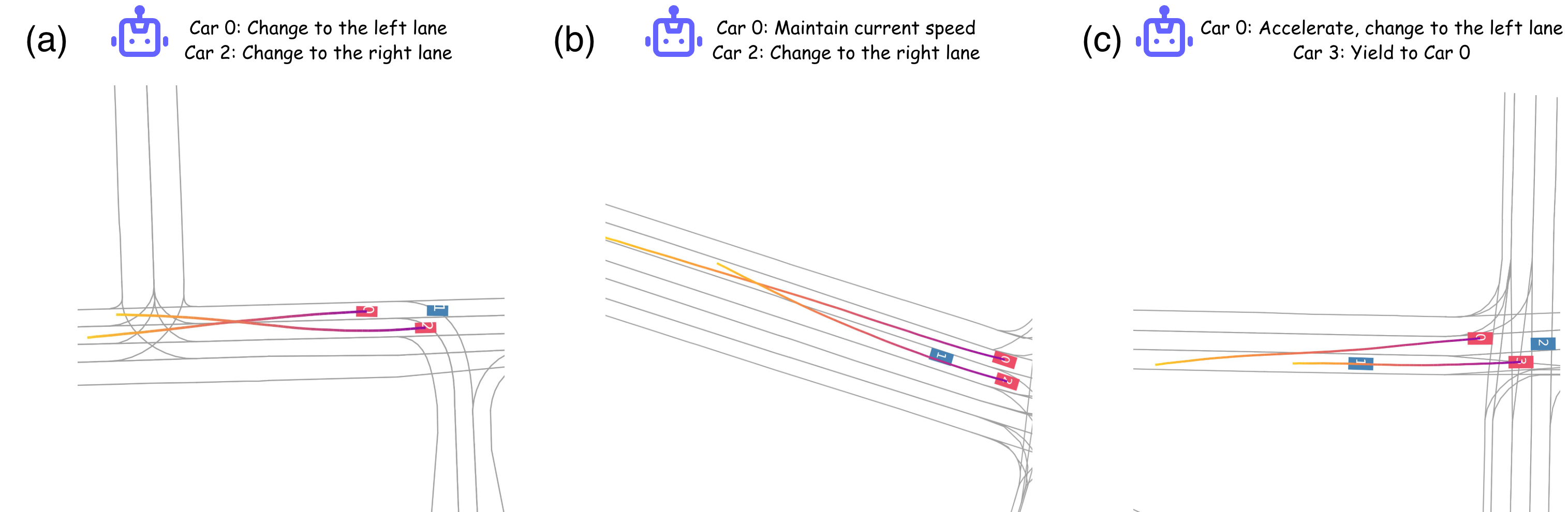}
    \caption{Results of interactive scenarios generated by \our. Our framework translates high-level VLM proposals into diverse, social interactive trajectories. Red vehicles correspond to the pair selected by the VLM for interaction.}
    \label{fig:socialdrivegen-example}
\end{figure}

\subsection{Qualitative Analysis: Interactive Scenario Generation}

Figure~\ref{fig:socialdrivegen-example} provides a qualitative analysis of scenarios generated by \our. Each subplot visualizes the trajectories of two interacting vehicles selected by the VLM, generated in response to high-level adversarial proposals. As illustrated, the generated trajectories are physically plausible, maintaining kinematic smoothness and collision-free states. Crucially, the results highlight the diversity of our framework, which successfully synthesizes a wide spectrum of driving behaviors—ranging from cooperative merging to competitive gaming.

Specifically, in Figure~\ref{fig:socialdrivegen-example}(a), both Car 0 and Car 2 attempt a concurrent lane change, resulting in a coordinated cooperative maneuver. In Figure~\ref{fig:socialdrivegen-example}(b), a non-cooperative scenario is synthesized where Car 0 maintains its speed and refuses to yield to the merging Car 2. In Figure~\ref{fig:socialdrivegen-example}(c), Car 3 exhibits yielding behavior, decelerating to accommodate Car 0’s lane change intent. Collectively, these examples demonstrate our framework's capability to translate high-level VLM proposals into concrete scenarios characterized by enhanced interaction intensity.

\subsection{Component Analysis and Ablation Study}

To validate the individual contributions of our core modules, we conduct an ablation study focusing on two critical aspects: the decision-making capability of the VLM planner and the optimization efficiency of the diffusion guidance mechanism. Table~\ref{tab:ablation_study} summarizes the performance in terms of Engagement Ratio and kinematic dynamics (Relative Velocity and Acceleration).

\begin{table}[ht]
\centering
\renewcommand{\arraystretch}{1.1}
\caption{Ablation Study on the Components of \our. We evaluate the contribution of the two stage VLM-based proposal generation and the stepwise guidance mechanism.}
\label{tab:ablation_study}
\setlength{\tabcolsep}{0pt}
\begin{tabular}{lccc}
\toprule
\textbf{Method} & 
\parbox[c]{3cm}{\centering \textbf{Engagement \\ Ratio }\\ ($\text{TTC} < 4.0$s)} & 
\parbox[c]{2.25cm}{\centering \textbf{Relative \\ Velocity \\ ($m/s$)}} & 
\parbox[c]{2.25cm}{\centering \textbf{ Acceleration \\ ($m/s^2$)}} \\
\midrule

 Random Proposal  & 34.55 & 3.01 & 6.50 \\
 Single Stage VLM & 50.91 & 3.65 & 7.25 \\
\midrule

    Diffusion-ES~\citep{yang2024diffusion} & 60.00 & 4.10 & 7.85 \\
\midrule

    \textbf{Ours (Full)} & \textbf{67.27} & \textbf{4.32} & \textbf{8.11} \\
\bottomrule
\end{tabular}
\end{table}

\paragraph{Effectiveness of the Two-Stage VLM Planner}

We first investigate the necessity of our hierarchical planning strategy by comparing it against two baselines: (1) \textit{Random Proposal}, which assigns pre-defined actions stochastically, and (2) \textit{Single-Stage VLM}, which attempts to generate proposals directly from the initial scene without intermediate reasoning.

As shown in Table~\ref{tab:ablation_study}, \our achieves an Engagement Ratio of $67.27\%$, surpassing the Single-Stage ($50.91\%$) and Random ($34.55\%$) methods by significant margins. The poor performance of the Random baseline highlights that high-stakes interactions are rare and require intelligent initiation. More importantly, the 16.36\% performance gap between our method and the Single-Stage VLM validates our hierarchical design philosophy. Complex traffic scenarios involve subtle dynamic cues that are difficult to capture via a direct mapping from raw states to proposals. By decomposing the task into a ``Describe-then-Propose'' pipeline, our framework mimics a Chain-of-Thought reasoning process. The first stage forces the model to explicitly articulate the global context and social intentions into a semantic description $D_{\text{scene}}$. This intermediate reasoning step serves as a critical grounding signal, enabling the second stage to strategically select the most relevant vehicle pair and formulate precise adversarial goals that are deeply contextualized rather than visually reactive.

\paragraph{Effectiveness of Iterative Guidance}

Second, we evaluate the impact of our gradient-free guidance mechanism. We compare our step-wise approach against \textit{Diffusion-ES}~\citep{yang2024diffusion}, a baseline that applies reward-based guidance only at the terminal step of the denoising process.

The results demonstrate that \our outperforms the baseline in both interaction success (higher Engagement Ratio) and dynamic intensity (higher Relative Velocity and Acceleration). We attribute this to the fundamental difference in optimization timing. The diffusion process generates trajectories by progressively refining noise; applying guidance only at the final step is often ``too late'' to significantly alter the trajectory's topological structure or correct early-stage deviations. in contrast, our iterative guidance actively optimizes the population distribution at each denoising step. This allows the model to correct the trajectory manifold early in the generation process, effectively steering the agents toward higher-reward solutions. Consequently, \our is capable of generating scenarios that are not only compliant with the proposal but also exhibit the challenging dynamics and high speeds required for robust safety testing.



\subsection{Controllability via Social Preferences}

To analyze the behavioral trade-offs induced by different social parameters, we shift our focus to the balance between interaction intensity and task performance. We report Engagement Ratio, Acceleration, and specifically the Extrinsic Reward, as the latter directly reflects the agent's adherence to navigational goals when social weights vary. Our proposed reward decomposition in Eq.~\eqref{eq:social_reward} introduces two interpretable control knobs: the \textbf{social orientation angle} $\phi$, which governs the agent's interaction style (e.g., aggressive vs. prosocial), and the \textbf{intrinsic reward magnitude} $\lambda$, which regulates the trade-off between social compliance (safety/comfort) and task completion (extrinsic goals). We set a rational egoist ($\lambda=1, \phi=0$) as the baseline. To demonstrate the controllability of our framework, we configure surrounding social vehicles as rational egoists to simulate a standard traffic environment.

\begin{table}[ht]
\centering
\renewcommand{\arraystretch}{1.1}
\caption{Quantitative comparison of generated trajectories under varying social preference parameters ($\lambda, \phi$), demonstrating their impact on interaction intensity, acceleration, and task performance.}
\label{tab:social_exp}
\setlength{\tabcolsep}{0pt}
\begin{tabular}{lccc}
\toprule
\parbox[c]{3.25cm}{\textbf{Social \\ Preference}} & 
\parbox[c]{3cm}{\centering \textbf{Engagement \\ Ratio }\\ ($\text{TTC} < 4.0$s)} & 
\parbox[c]{3cm}{\centering \textbf{Acceleration \\ ($m/s^2$)}} & 
\parbox[c]{3cm}{\centering \textbf{ Extrinsic \\ Reward}} \\
\midrule

$\lambda = 1.0, \phi = 0$  & 67.27 & 8.11 & 0.37 \\
\midrule
$\lambda = 1.0, \bm{\phi = \frac{\pi}{4}}$ & 63.64 (\textcolor{softblue}{-5.40\%}) & 7.97 (\textcolor{softblue}{-1.73\%}) & 0.48 (\textcolor{softred}{+29.73\%}) \\
$\lambda = 1.0, \bm{\phi = -\frac{\pi}{4}}$ & 72.73 (\textcolor{softred}{+8.12\%}) & 8.16 (\textcolor{softred}{+0.62\%}) & 0.31 (\textcolor{softblue}{-16.22\%}) \\
$\bm{\lambda = 0.5}, \phi = 0$ & 65.45 (\textcolor{softblue}{-2.71\%}) & 8.17 (\textcolor{softred}{+0.74\%}) & 0.60 (\textcolor{softred}{+62.16\%}) \\
$\bm{\lambda = 0.3}, \phi = 0$ & 69.09 (\textcolor{softred}{+2.71\%}) & 8.28 (\textcolor{softred}{+2.10\%}) & 0.69 (\textcolor{softred}{+86.49\%}) \\
\bottomrule
\end{tabular}
\end{table}

\paragraph{Impact of Social Orientation ($\phi$)}

We first examine how the directional preference $\phi$ influences the style of interaction. By modulating $\phi$ with fixed $\lambda$, we effectively shift the agent's behavior along the SVO spectrum without altering its high-level goal. As shown in Table~\ref{tab:social_exp}, setting $\phi = -\frac{\pi}{4}$ (representing a competitive or adversarial persona) leads to an $8.12\%$ increase in Engagement Ratio compared to the baseline ($\phi=0$). This suggests that negative angular preferences encourage the agent to prioritize its own intrinsic utility over the opponent's, resulting in more assertive behaviors such as forcing a merge or refusing to yield, thereby intensifying the interaction. Conversely, a prosocial setting ($\phi = \frac{\pi}{4}$) reduces the Engagement Ratio by $5.40\%$, as the agent actively seeks to minimize the opponent's discomfort, often yielding early to avoid high-stakes scenarios. This validates that $\phi$ serves as an effective parameter for controlling the ``aggressiveness'' of the generated scenarios.

\paragraph{Impact of Intrinsic Magnitude ($\lambda$)}

Next, we evaluate the role of $\lambda$ in balancing social constraints against task objectives. In our framework, a high $\lambda$ implies a socially compliant driver who strictly adheres to safety and comfort norms, potentially at the cost of task efficiency. As we progressively decrease $\lambda$ from $1.0$ to $0.3$, the agent becomes increasingly goal-oriented and egoistic, reducing the weight of social penalties. The results in Table~\ref{tab:social_exp} reveal a dramatic monotonic increase in Extrinsic Reward, peaking at $+86.49\%$ for $\lambda=0.3$. This indicates that as social constraints are relaxed, the agent becomes more efficient at achieving the VLM-specified objectives (e.g., reaching a target lane quickly), even if it requires more dynamic maneuvers. Notably, the Engagement Ratio remains high ($69.09\%$) at lower $\lambda$, confirming that the agent maintains interactivity but changes its motivation: from ``interacting carefully'' to ``interacting efficiently''. This validates our decomposition strategy, proving that we can disentangle extrinsic task motivations from intrinsic social constraints to generate diverse driving profiles.

\section{Conclusion}
We introduced \our, a hierarchical framework that pairs a two-stage VLM proposal generation for proposing high-risk scenarios with reward-guided diffusion model for generating them. Our framework integrates reward models derived from social psychology to controllably generate the egoism--altruism spectrum of driving behaviors. This work provides a scalable and controllable pathway for populating the ``long-tail" of autonomous driving, opening a new avenue for robustly training and testing agents against the full spectrum of complex, safety-critical social interactions.
\bibliographystyle{rlc}
\bibliography{ref}

\end{document}